\def\al{\alpha}
\def\ma{\mathcal}
\def\ra{\rangle}
\def\la{\langle}
\def\be{\begin{equation}}
\def\ee{\end{equation}}
\def\ba{\begin{array}}
\def\ea{\end{array}}

\documentclass[12pt]{article}
\usepackage{cite}
\usepackage{CJK}
\usepackage{amssymb}
\usepackage{amsmath}
\usepackage{enumerate}
\usepackage{graphicx}
\usepackage{amsfonts}
\usepackage{url}
\usepackage{bm}
\usepackage{pifont}
\usepackage{epsfig,subfigure,dsfont,amsthm,amsbsy,mathrsfs,amscd}
\newtheorem{theorem}{Theorem}
\newtheorem{lemma}{Lemma}

\input amssym.def

\begin{document}
\title{\large\bf On Local Unitary Equivalence of Two and Three-qubit States}
\author{Bao-zhi Sun$^{1}$ $^\ast$\& Shao-Ming Fei$^{2}$ \& Zhi-xi Wang$^{2}$ \\[10pt]
\footnotesize
\small $^{1}$School of Mathematics, Qufu Normal University, JiNing, Shandong 273165, China\\
\footnotesize
\small $^{2}$School of Mathematical Sciences, Capital Normal University, Beijing 100048, China}
\date{}

\maketitle

\centerline{$^\ast$ Correspondence to sun\_bao\_zhi@163.com}
\bigskip

\begin{abstract}

We study the local unitary equivalence for two and three-qubit mixed states by investigating
the invariants under local unitary transformations.
For two-qubit system, we prove that the determination of the local unitary
equivalence of 2-qubits states only needs 14 or less invariants for arbitrary two-qubit states.
Using the same method, we construct invariants for three-qubit mixed states.
 We prove that these invariants are sufficient to guarantee the LU equivalence of certain
kind of three-qubit states. Also, we make a comparison with earlier works.

\end{abstract}
\bigskip

Nonlocality is one of the astonishing phenomena in quantum mechanics. It is not only important in philosophical considerations of the nature of quantum theory, but also the key ingredient in quantum computation and communications such as cryptography \cite{nielsen}. From the point of view of nonlocality, two states are completely equivalent if one can be transformed into the other by means of local unitary (LU) transformations. Many crucial properties such as the degree of entanglement \cite{eof1,eof2}, maximal violations of Bell inequalities \cite{bell1,bell2,bell3,bell4} and the teleportation fidelity \cite{tel1,tel2} remain invariant under LU transformations. For this reason, it has been a key problem to determine whether or not two states are LU equivalent.

There have been a plenty of results on invariants under LU transformations \cite{makhlin,3-qubit, gm, bliu, Li2, Ma, jpa, JLLZF, Sp,zhou,JF1}. However, one still does not have a complete set of such LU invariants which can operationally determine the LU equivalence of any two states both necessarily and sufficiently, except for 2-qubit states and some special 3-qubit states.
For the 2-qubit state case, Makhlin presented a set of 18 polynomial LU invariants in \cite{makhlin}.
In \cite{JF1} the authors constructed a set of very simple invariants which are less than the ones constructed in \cite{makhlin}.
Nevertheless, the conclusions are valid only for special (generic) two-qubit states and an error
occurred in the proof.
In this paper, we corrected the error in \cite{JF1} by adding some missed invariants, and prove that the determination of the local unitary equivalence of 2-qubits states only needs 14 or less invariants for arbitrary two-qubit states.
Moreover, we prove that the invariants in \cite{JF1} plus some invariants from
triple scalar products of certain vectors are complete for a kind of 3-qubit states.

\medskip
\noindent{\bf Results}
\medskip

A general $2$-qubit state can be expressed as:
$$
\rho=\frac{1}{4}(I_2\otimes I_2+\sum_{i=1}^3T_1^i\sigma_i\otimes I_2+\sum_{j=1}^3T_2^jI_2\otimes\sigma_j+\sum_{i,j=1}^3T_{12}^{ij}\sigma_i\otimes\sigma_j),
$$
where $I$ is the $2\times2$ identity matrix, $\sigma_{i}$, $i=1,2,3$, are Pauli matrices and $T_1^i=\mathrm{tr}(\rho(\sigma_i\otimes I))$ etc.
Two two-qubit states $\rho$ and
$$
\hat{\rho}=\frac{1}{4}(I_2\otimes I_2+\sum_{i=1}^3\hat{T}_1^i\sigma_i\otimes I_2+\sum_{j=1}^3\hat{T}_2^jI_2\otimes\sigma_j+\sum_{i,j=1}^3\hat{T}_{12}^{ij}\sigma_i\otimes\sigma_j)
$$
are called LU equivalent if there are some $U_i\in U(2)$, $i=1,2$, such that
$\hat{\rho}=(U_1\otimes U_2)\rho(U_1^\dag\otimes U_2^\dag)$.
By using the well-known double-covering map $SU(2)\longrightarrow SO(3)$, one has that for all $U\in SU(2)$,
there is a matrix $O=(o_{kl})\in SO(3)$, such that $U\sigma_kU^\dag=\sum_{l=1}^3o_{kl}\,\sigma_l$. Therefore,
$\rho$ and $\hat{\rho}$ are LU equivalent if and only if there are some $O_i\in SO(3)$, $i=1,2$, such that
\begin{equation}\label{condition-2}\begin{array}{l}\hat{T}_1=O_1T_1,\ \ \ \ \hat{T}_2=O_2T_2,\\[1mm]
\hat{T}_{12}=O_1T_{12}O_2^t.
\end{array}\end{equation}
One has two sets of vectors,
\begin{equation}\label{sets}
\begin{array}{l}
S_1=\{T_1,T_{12}T_2,T_{12}T_{12}^tT_1,T_{12}T_{12}^tT_{12}T_2,\cdots\},\\[2mm]
S_2=\{T_2,T_{12}^tT_1,T_{12}^tT_{12}T_2,T_{12}^tT_{12}T_{12}^tT_1,\cdots\}.
\end{array}
\end{equation}
For convenience, we denote $S_1=\{\mu_i|i=1,2,\cdots\}$, $S_2=\{\nu_i|i=1,2,\cdots\}$, i.e., $\mu_1=T_1$, $\mu_2=T_{12}T_2$, $\mu_3=T_{12}T_{12}^tT_1$ and so on.
The vectors  $\mu_i$ ($\nu_i$)
are transformed into $O_1\mu_i$ ($O_2\nu_i$) under local unitary transformations. Otherwise, local unitary transformation can transform $\mu_i\times\mu_j$ to $O_1(\mu_i\times\mu_j)$ and $\nu_i\times\nu_j$ to $O_2(\nu_i\times\nu_j)$.
Hence it is direct to verify that the inner products $\langle\mu_i,\mu_j\rangle$, $\langle\nu_i,\nu_j\rangle$, $i,j=1,2,\cdots$, and $(\mu_i,\mu_j,\mu_k)\equiv\langle\mu_i,\mu_j\times\mu_k\rangle$, $(\nu_i,\nu_j,\nu_k)\equiv\langle\nu_i,\nu_j\times\nu_k\rangle$, $i,j,k=1,2,\cdots$, are invariants under
local unitary transformations. Moreover, from the transformation ${T}_{12}\to O_1T_{12}O_2^t$, we have
that $\mathrm{tr}(T_{12}T_{12}^t)^\alpha$, $\alpha=1,2,\cdots,$ and $\det T_{12}$ are also LU invariants.

For a set of 3-dimensional real vectors $S=\{\mu_i|i=1,2,\cdots\}$, we denote $\dim\la S\ra$
the dimension of the real linear space spanned by $\{\mu_i\}$, i.e., the number of linearly
independent vectors of $\{\mu_i\}$.
As the vectors in $S_1$ and $S_2$ are three-dimensional, there are at most 3
linearly independent vectors in each vector sets $S_1$ and $S_2$.

First note that, given two sets of 3-dimensional real vectors
$S=\{\mu_i|i=1,2,\cdots\}$ and $\hat{S}=\{\hat{\mu}_i|i=1,2,\cdots\}$, if the inner products $\la\mu_i,\mu_j\ra=\la\hat{\mu}_i,\hat{\mu}_j\ra$, then the following conclusions are true:
(i) $\dim\la S\ra=\dim\la\hat{S}\ra$;
(ii) The corresponding subsets of $S$ and $\hat{S}$
have the same linear relations;
(iii) There exist $O\in O(3)$ such that $\hat{\mu}_i=O\mu_i$. Furthermore, using $(\mu_i,\mu_j,\mu_k)=(\hat{\mu}_i,\hat{\mu}_j,\hat{\mu}_k)$, we can get that $O\in SO(3)$.
If $\dim\la S\ra=3$, then $O$ is unique. For $\dim\la S\ra<3$, $(\mu_i,\mu_j,\mu_k)=(\hat{\mu}_i,\hat{\mu}_j,\hat{\mu}_k)=0$, and there is at least one $O\in SO(3)$ such that $\hat{\mu}_i=O\mu_i$.

Next we clarify the independent invariants in $S_1$ and $S_2$.
From the definition of $\mu_i,\nu_i$, we have
$$\langle\mu_i,\mu_j\rangle=\left\{\begin{array}{cl}T_1^t(T_{12}T_{12}^t)^{a_{ij}}T_1,& \mathrm{if}\   i,j\ \mathrm{are\ odd}\\
T_2^t(T_{12}^tT_{12})^{a_{ij}}T_2,& \mathrm{if}\ i,j\ \mathrm{are\ even}\\
T_1^t(T_{12}T_{12}^t)^{b_{ij}}T_{12}T_2,&\mathrm{if}\ i+j\ \mathrm{is\ odd}\\
\end{array}\right.$$
$$\langle\nu_i,\nu_j\rangle=\left\{\begin{array}{cl}T_2^t(T_{12}^tT_{12})^{a_{ij}}T_2,& \mathrm{if}\   i,j\ \mathrm{are\ odd}\\
T_2^1(T_{12}T_{12}^t)^{a_{ij}}T_1,& \mathrm{if}\ i,j\ \mathrm{are\ even}\\
T_1^t(T_{12}T_{12}^t)^{b_{ij}}T_{12}T_2,&\mathrm{if}\ i+j\ \mathrm{is\ odd}\\
\end{array}\right.$$
where $a_{ij}=(i+j-2)/2$, $b_{ij}=(i+j-3)/2$.
From Hamilton-Cayley theorem, when $a_{ij},b_{ij}\geq 3$, the invariants $\la\mu_i,\mu_j\ra$ and $\la\nu_i,\nu_j\ra$ can be linearly represented by $\la\mu_p,\mu_q\ra$, $\la\nu_p,\nu_q\ra$,
$a_{pq}, b_{pq}<3$. Therefore there are only  9 linearly independent invariants:
$\la\mu_i,\mu_i\ra$, $\la\nu_i,\nu_i\ra$, $i=1,2,3$, and $\la\mu_1,\mu_j\ra$, $j=2,4,6.$
We denote them as $L=\{\la\mu_i,\mu_i\ra, \la\nu_i,\nu_i\ra, \la\mu_1,\mu_j\ra|i=1,2,3, j=2,4,6\}.$

For 2-qubit states $\rho$ and $\hat{\rho}$, if $\dim \la S_1\ra=\dim \la\hat{S}_1\ra=3$, we need one more invariant $(\mu_{r_0},\mu_{s_0},\mu_{t_0})$ to guarantee that there is an $O_1\in SO(3)$ such that $O_1\mu_i=\hat{\mu}_i$,  for any $i$. Here $\mu_{r_0},\mu_{s_0}$ and $\mu_{t_0}$ are arbitrary three linear independent vectors in $S_1$. If $\dim \la S_1\ra=\dim \la\hat{S}_1\ra<3$, then the invariants in $L$ are enough to guarantee the existence of $O_1$. Similar conclusions are true for $S_2$ and $\hat{S}_2$.

Let $\mu_{r_0},\mu_{s_0}$ and $\mu_{t_0}$ ($\nu_{r_0},\nu_{s_0}$ and $\nu_{t_0}$) denote arbitrary three linear independent vectors in $S_1$ ($S_2$) if $\dim\la S_1\ra=3$ ($\dim\la S_2\ra=3$).
For the case that at least one  of $\dim\la S_1\ra$ and $\dim\la S_2\ra$ is 3, we have

\begin{theorem} Two 2-qubit states are LU equivalent if and only if they have same values of the invariants in $L$, the invariant $(\mu_{r_0},\mu_{s_0},\mu_{t_0})$ and/or $(\nu_{r_0},\nu_{s_0},\nu_{t_0})$ if $\dim\la S_1\ra=3$ and/or $\dim\la S_2\ra=3$.
\end{theorem}

See Methods for the proof of Theorem 1.

For the case both  $\dim\langle S_1\rangle<3$ and $\dim\langle S_2\rangle<3$, we also have $O_2T_{12}^t\mu_i=\hat{T}_{12}^tO_1\mu_i$ for some $O_i\in SO(3)$. But this
does not necessarily give rise to $\hat{T}_{12}=O_1T_{12}O_2^t$.
In order to discuss these cases, we need the following result.

\begin{lemma}\label{correlation}
For two-qubit states $\rho$ and $\hat{\rho}$, if $\mathrm{tr}(T_{12}T_{12}^t)^\alpha=\mathrm{tr}(\hat{T}_{12}\hat{T}_{12}^t)^\alpha$, $\alpha=1,2$ and $\det T_{12}=\det\hat{T}_{12}$.
then $\hat{T}_{12}=O_1T_{12}O_2^t$ for some $O_1,O_2\in SO(3)$.
\end{lemma}

See Methods for the proof of Lemma 1.

For the completeness of the set of invariants, we also need an extra invariant $\mathbf{I}=\epsilon_{ijk}\epsilon_{lmn}T_1^iT_2^lT_{12}^{jm}T_{12}^{kn}$, here $\epsilon_{ijk}$ and $\epsilon_{lmn}$ are Levi-Cevita symbol. Now we discuss the case of $\dim S_i=\dim\hat{S}_i<3$, $i=1,2$.

\begin{theorem}
Two 2-qubit states with $\dim S_i=\dim\hat{S}_i<3$, $i=1,2$ are local unitary equivalent if and only if they have the same values of the invariants in $L$, and the invariants
$\mathrm{tr}(T_{12}T_{12}^t)^\alpha$, $\alpha=1,2$, $\det T_{12}$ and \textbf{I}.
\end{theorem}

See Methods for the proof of Theorem 2.

From Theorem 1 and 2 we see that for the case at least one of $\la S_i\ra$ has dimension three,
we only need 11 or 10 invariants to determine the local unitary equivalence of two 2-qubit states:
namely, 9 invariants from $L$, and $(\mu_{r_0},\mu_{s_0},\mu_{t_0})$ and/or $(\nu_{r_0},\nu_{s_0},\nu_{t_0})$. If both the dimensions of $\la S_1\ra$ and $\la S_2\ra$ are less than 3,
then $(\mu_{r_0},\mu_{s_0},\mu_{t_0})=(\nu_{r_0},\nu_{s_0},\nu_{t_0})=0$. To determine the LU equivalence, we need invariants from $L$, $\textbf{I}$, $\mathrm{tr}(T_{12}T_{12}^t)^\alpha$, $\alpha=1,2$, and $\det T_{12}$.
Hence we need at most 13 independent invariants.  In \cite{JF1}, the authors considered only the generic case of $\dim\langle S_i\rangle=3,$  $i=1$ and $2$, in which the important invariants $(\mu_{r_0},\mu_{s_0},\mu_{t_0})$ and $(\nu_{r_0},\nu_{s_0},\nu_{t_0})$ are missed. By adding these missed invariants, we have remedied the error in \cite{JF1} and, moreover, generalized the method to the case of $\dim\langle S_i\rangle=3$ for $i=1$ or $2$.

As an example, let we consider the states $\rho$ and $\hat{\rho}$ with $T_1=(1,1,1)^t$ and $\hat{T}_1=(1,1,-1)^t$, respectively. $T_{12}$ and $\hat{T}_{12}$ have the same singular values
that are all different. Hence $\dim\langle S_1\rangle=\dim\langle \hat{S}_1\rangle=3$. In this case the invariants from \cite{JF1} have the same values for $\rho$ and $\hat{\rho}$. Nevertheless,
taking $\mu_{r_0}=T_1$, $\mu_{s_0}=T_1T_{12}T_{12}^t$ and $\mu_{t_0}=T_1(T_{12}T_{12}^t)^2$, and correspondingly, $\hat{\mu}_{r_0}=\hat{T}_1$, $\hat{\mu}_{s_0}=\hat{T}_1\hat{T}_{12}\hat{T}_{12}^t$, and $\hat{\mu}_{t_0}=\hat{T}_1(\hat{T}_{12}\hat{T}_{12}^t)^2$, we find that the triple scalar invariant we added are different for $\rho$ and $\hat{\rho}$, $(\mu_{r_0}, \mu_{s_0}, \mu_{t_0})=-(\hat{\mu}_{r_0}, \hat{\mu}_{s_0}, \hat{\mu}_{t_0})\neq 0$. Therefore, $\rho$ and $\hat{\rho}$ are not locally equivalent.

The expression of a complete set of LU invariants depends on the form of the invariants.
Different constructions of LU invariants may give different numbers of the invariants in the complete set,
and may have different advantages. Obviously the eigenvalues of
a density matrix are LU invariants. Based on the eigenstate decompositions of density matrices,
in \cite{zhou} complete set of LU invariants are presented for arbitrary dimensional bipartite states.
Nevertheless, such kind of construction of invariants results in problems
when the density matrices are degenerate, i.e. different eigenstates have the same eigenvalues.
The 18 LU invariants constructed in \cite{makhlin} are based on the Bloch representations
of 2-qubit states and has no such problem as in \cite{zhou}.
However, these 18 invariants are complete but more than necessary in the sense that
the number of independent invariants can be reduced by suitable constructions of the invariants.
The LU invariants constructed in \cite{JF1} are also in terms of Bloch representations.
Such constructed invariants work for both non-degenerate and degenerate states.
Nevertheless, the invariants: $\mathbf{I}$, $(\mu_{r_0},\mu_{s_0},\mu_{t_0})$,
$(\nu_{r_0},\nu_{s_0},\nu_{t_0})$ and $\det T_{12}=\det\hat{T}_{12}$ make the corresponding
theorems  incorrect even for generic cases studied in \cite{JF1}.
By adding these invariants, our set of invariants work for arbitrary 2-qubit states.
In fact, a set of complete LU invariants characterizes completely the LU orbits in the quantum state space.
Generally such orbits are not manifolds, but varieties. For example, the
set of pure states is a symplectic variety \cite{Ref}. For general mixed states, the situation
is much more complicated \cite{JPA}. Our results would highlight the analysis on the structures of LU orbits.


Now we come to discuss the case of three-qubit system.
A three-qubit state $\rho$ can be written as:
$$
\begin{array}{rl}\rho&=\displaystyle\frac{1}{8}(I_2\otimes I_2\otimes I_2+\sum_{i=1}^3T_1^i\sigma_i\otimes I_2\otimes I_2+\sum_{j=1}^3T_2^jI_2\otimes\sigma_j\otimes I_2+\sum_{k=1}^3T_3^kI_2\otimes I_2\otimes\sigma_k\\[1mm]
&+\displaystyle\sum_{i,j=1}^3T_{12}^{ij}\sigma_i\otimes\sigma_j\otimes I_2+\sum_{i,k=1}^3T_{13}^{ik}\sigma_i\otimes  I_2\otimes \sigma_k +\sum_{j,k=1}^3T_{12}^{jk}I_2\otimes\sigma_j\otimes\sigma_k\\[1mm]
&+\displaystyle\sum_{i,j,k=1}^3T_{123}^{ijk}\sigma_i\otimes\sigma_j\otimes \sigma_k).
\end{array}
$$
One has the coefficient vectors $T_1,T_2,T_3$, coefficient matrices $T_{12},T_{23},T_{13}$ and coefficient tensor $T_{123}$.
Now, $\rho$ and $\hat{\rho}$ are LU equivalent if and only if there are $O_i\in SO(3)$, $i=1,2,3$, such that
$\hat{T}_i=O_iT_i$, $\hat{T}_{ij}=O_i\otimes O_jT_{ij}$, $\hat{T}_{123}=O_1\otimes O_2\otimes O_3 T_{123}$.
For simplicity we denote $t_{ijk}\equiv T_{123}^{ijk}$ and
$$\ba{l}
T_{1|23}=\left(\ba{ccccccccc}
t_{111}&t_{112}&t_{113}&t_{121}&t_{122}&t_{123}&t_{131}&t_{132}&t_{133}\\
t_{211}&t_{212}&t_{213}&t_{221}&t_{222}&t_{223}&t_{231}&t_{232}&t_{233}\\
t_{311}&t_{312}&t_{313}&t_{321}&t_{322}&t_{323}&t_{331}&t_{332}&t_{333}
\ea\right),\\[10mm]
T_{2|13}=\left(\ba{ccccccccc}
t_{111}&t_{112}&t_{113}&t_{211}&t_{212}&t_{213}&t_{311}&t_{312}&t_{313}\\
t_{121}&t_{122}&t_{123}&t_{221}&t_{222}&t_{223}&t_{321}&t_{322}&t_{323}\\
t_{131}&t_{132}&t_{133}&t_{231}&t_{232}&t_{233}&t_{331}&t_{332}&t_{333}
\ea\right),\\[10mm]
T_{3|12}=\left(\ba{ccccccccc}
t_{111}&t_{121}&t_{131}&t_{211}&t_{221}&t_{231}&t_{311}&t_{321}&t_{331}\\
t_{112}&t_{122}&t_{132}&t_{212}&t_{222}&t_{232}&t_{312}&t_{322}&t_{332}\\
t_{113}&t_{123}&t_{133}&t_{213}&t_{223}&t_{233}&t_{313}&t_{323}&t_{333}
\ea\right).
\ea
$$
Also, we write $\ma{T}_1=T_{1|23}T_{1|23}^t,\  \ma{T}_2=T_{2|13}T_{2|13}^t,\  \ma{T}_3=T_{3|12}T_{3|12}^t$ and
$\ma{T}_{23}=T_{1|23}^tT_{1|23},\  \ma{T}_{13}=T_{2|13}^tT_{2|13},\  \ma{T}_{12}=T_{3|12}^tT_{3|12}$.
Similar to to the two-qubit case, one has three sets of vectors,
$$
\ba{l}
S_1=\{\ma{T}_1^{r-1}T_1,\ma{T}_1^{r-1}T_{12}\ast\ast, \ma{T}_1^{r-1}T_{13}\ast\ast, \ma{T}_1^{r-1}T_{1|23}\ast\ast\},\\[1mm]
S_2=\{\ma{T}_2^{r-1}T_2,\ma{T}_2^{r-1}T_{12}^t\ast\ast, \ma{T}_2^{r-1}T_{23}\ast\ast, \ma{T}_2^{r-1}T_{2|13}\ast\ast\},\\[1mm]
S_3=\{\ma{T}_3^{r-1}T_3,\ma{T}_3^{r-1}T_{13}^t\ast\ast, \ma{T}_3^{r-1}T_{23}^t\ast\ast, \ma{T}_3^{r-1}T_{3|12}\ast\ast\},
\ea
$$
where $r=1,2,3$ and $\ast\ast$ represents all the suitable vectors constructed from $T_{ij}$, $T_{i|jk}$, $\ma{T}_i$ and $T_i$ such that
the vectors in $S_i$ are transformed into $O_i S_i$ under LU transformations. For instance,
we have $T_{12}^tS_1\subset S_2$, $T_{13}^tS_1\subset S_3$, $T_{1|23}\,S_2\otimes S_3\subset S_1$ and so on,
where for $S_2=\{\nu_i|i=1,2,\cdots\}$ and $S_3=\{\omega_j|j=1,2,\cdots\}$, we have denoted $S_2\otimes S_3=\{\nu_i\otimes\omega_j|i,j=1,2,\cdots\}$ etc.
Because the vectors in $S_i$ are all 3-dimensional, we have $\dim\la S_i\ra\leq 3$.
The inner products $\la\mu_i,\mu_j\ra$, $\la\nu_i,\nu_j\ra$ and $\la\omega_i,\omega_j\ra$, $i,j=1,2,\cdots$, are all invariants under LU transformations.
Using the method in \cite{JF1}, we now prove that these invariants together with the additional ones in theorem 3 are sufficient to guarantee the LU equivalence of certain kind of three-qubit states with at least two of $\dim\la S_i\ra=3$ for $i=1,2,3$.

 \begin{theorem}\label{3-qubits}
 Given two 3-qubit states $\rho$ and $\hat{\rho}$, if $\la X_i,X_j\ra=\la\hat{X}_i,\hat{X}_j\ra$,
 $(X_i,X_j,X_k)=(\hat{X}_i,\hat{X}_j,\hat{X}_k)$ for $X=\mu,\nu,\omega$ and $i,j,k=1,2,\cdots$,
 and $\dim\la S_i\ra=\dim\la \hat{S}_i\ra=3$ for at least two $i\in{1,2,3}$,
 then $\rho$ and $\hat{\rho}$ are LU equivalent.
 \end{theorem}

See Methods for the proof of Theorem 3.

If at most one of $\dim\la S_i\ra$ is 3, things become more complicated.
Now we give a comparison with the results in \cite{3-qubit}.
For 3-qubit states $\rho$ and $\hat{\rho}$, if
\be\label{0-T}
\mathrm{tr}(\ma{T}_i^r)=\mathrm{tr}(\hat{\ma{T}}_i^r),~~~
 T_i^t\ma{T}_i^{r-1}T_i=\hat{T}_i^t\hat{\ma{T}}_i^{r-1}\hat{T}_i,~~~~  r,i=1,2,3,
\ee
then there are $P_i,\hat{P}_i\in O(3)$ such that
\be\label{total}P_i\ma{T}_iP_i^t=\left(\ba{ccc}t_{i1}&&\\ &t_{i2}&\\ &&t_{i3}\ea\right)=\hat{P}_i\hat{\ma{T}}_i\hat{P}_i^t,~~~ P_iT_i=\hat{P}_i\hat{T}_i=\left(\ba{c}a_{i1}\\ a_{i2}\\ a_{i3}\ea\right).
\ee
Denote
$$Y_i\equiv\left(\ba{ccc}a_{i1}&a_{i2}&a_{i3}\\ t_{i1}a_{i1}&t_{i2}a_{i2}&t_{i3}a_{i3}\\  t_{i1}^2a_{i1}&t_{i2}^2a_{i2}&t_{i3}^2a_{i3}\ea\right)
=\left(\ba{ccc}1&1&1\\ t_{i1}&t_{i2}&t_{i3}\\  t_{i1}^2&t_{i2}^2&t_{i3}^2\ea\right)
\left(\ba{ccc}a_{i1}&&\\ &a_{i2}&\\ &&a_{i3}\ea\right)\equiv\Lambda_i\Theta_i.
$$
The results in \cite{3-qubit} concluded that  $\rho$ and $\hat{\rho}$ are local unitary equivalent if and only if the invariants in Theorem \ref{3-qubits}, together with the invariants $\mathrm{tr}(\ma{T}_i^r)$, $ r,i=1,2,3$ for the case of $\det\Lambda_i\Theta_i\neq0$, $i=1,2,3$.
Obviously, if $\det\Lambda_i\Theta_i\neq0$, $P_iT_i$, $P_i\ma{T}_iT_i$ and $P_i\ma{T}_i^2T_i$ are linear independent, so all $\dim\la S_i\ra=3$.
But $\dim\la S_i\ra=3$ does not necessarily imply  $\det\Lambda_i\Theta_i\neq 0$. Here we only need that two of the $\dim\la S_i\ra$ are 3.
So we  give the sufficient conditions for local unitary equivalence
of more states than the ones given in \cite{3-qubit}.

\medskip
\noindent{\bf Conclusion}

We study the local unitary equivalence for two and three-qubit mixed states by
investigating the invariants under local unitary transformations.
We corrected the error in [20] by adding some missed invariants, and prove that
the determination of the local unitary equivalence of 2-qubits states only needs 14 or less
invariants for arbitrary two-qubit states. Moreover, we prove that the invariants in [20] plus
some invariants from triple scalar products of certain vectors are complete for a kind of
3-qubit states.
Comparing with the results in \cite{3-qubit}, it has been shown that we judge the LU equivalence for a larger class of 3-qubit states.

\medskip
\noindent{\bf Methods}
\medskip

\medskip
\noindent{\bf Proof of Theorem 1}
 Suppose $\dim\langle S_1\rangle=\dim\langle\hat{S}_1\rangle=3$. From the construction of $S_1$ and
$S_2$, we have that $\nu_{i+1}=T_{12}^t\mu_i$, $\hat{\nu}_{i+1}=\hat{T}_{12}^t\hat{\mu}_i$, $i=1,2,\cdots$.
Then $O_2T_{12}^t\mu_i=O_2\nu_{i+1}=\hat{\nu}_{i+1}=\hat{T}_{12}^t\hat{\mu}_i=\hat{T}_{12}^tO_1\mu_i$, $i=1,2,\cdots$.
Since $\mu_{r_0},\mu_{s_0}$ and $\mu_{t_0}$ are linearly independent, $\det(\mu_{r_0}\ \mu_{s_0}\ \mu_{t_0})\neq 0$, where $(\mu_{r_0}\ \mu_{s_0}\ \mu_{t_0})$ denotes the
$3\times 3$ matrix given by the three column vectors $\mu_{r_0},\mu_{s_0}$ and $\mu_{t_0}$.
From $O_2T_{12}^t(\mu_{r_0}\ \mu_{s_0}\ \mu_{t_0})=\hat{T}_{12}^tO_1(\mu_{r_0}\ \mu_{s_0}\ \mu_{t_0})$, we get $O_2T_{12}^t=\hat{T}_{12}^t O_1$. Then $\hat{T}_{12}=O_1T_{12}O_2^t$.
The same result can be obtained from $\dim\langle S_2\rangle=\dim\langle\hat{S}_2\rangle=3$.
\hfill \rule{1ex}{1ex}

 \medskip
\noindent{\bf Proof of Lemma 1}
  From $\mathrm{tr}(T_{12}T_{12}^t)^\alpha=\mathrm{tr}(\hat{T}_{12}\hat{T}_{12}^t)^\alpha$, $\alpha=1,2$ and $\det T_{12}=\det\hat{T}_{12}$, one has that $T_{12}$ and $\hat{T}_{12}$ have the same singular values. According to the singular value decomposition, there are $P_i,\hat{P}_i\in O(3)$, $i=1,2$, such that $P_1T_{12}P_2^t=\hat{P}_1\hat{T}_{12}\hat{P}_2^t=diag(t_1,t_2,t_3)$, where $t_1,t_2$ and $t_3$ are the singular values. Set $O_1=\hat{P}_1^tP_1$, $O_2=\hat{P}_2^tP_2\in O(3)$, we have $\hat{T}_{12}=O_1T_{12}O_2^t$. From $\det T_{12}=\det\hat{T}_{12}$, we have that $\det O_1=\det O_2=\pm1$. If $\det O_1=\det O_2=-1$, we may change $P_i$ to $-P_i$ to have $O_i\in SO(3)$.
\hfill \rule{1ex}{1ex}

\medskip
\noindent{\bf Proof of Theorem 2}
 We only need to prove the ``only if" part, i.e., to find $O_1,O_2\in SO(3)$ such that $\hat{T}_{12}=O_1T_{12}O_2^t$, $\hat{T}_1=O_1T_{1}$, and $\hat{T}_2=O_2T_{2}$ for two 2-qubit states
$\rho$ and $\hat{\rho}$. From Lemma \ref{correlation}, we have $P_i,\hat{P}_i\in O(3)$, such that $\hat{P}_i^tP_i\in SO(3)$ and
\be\label{sd}
P_1T_{12}P_2^t=\hat{P}_1\hat{T}_{12}\hat{P}_2^t=\mathrm{diag}(t_1,t_2,t_3).
\ee
Hence
$$
P_1T_{12}T_{12}^tP_1^t=\hat{P}_1\hat{T}_{12}\hat{T}_{12}^t\hat{P}_1^t=P_2T_{12}^tT_{12}P_2^t
  =\hat{P}_2\hat{T}_{12}^t\hat{T}_{12}\hat{P}_2^t =\mathrm{diag}(t_1^2,t_2^2,t_3^2).
$$
Let $D=\mathrm{diag}(t_1,t_2,t_3)$, then $P_1S_1=\{P_1T_1,DP_2T_2,D^2P_1T_1,D^3P_2T_2,D^4P_1T_1,\cdots\}$,
$P_2S_2=\{P_2T_2, DP_1T_1,D^2P_2T_2,D^3P_1T_1,D^4P_2T_2,\cdots\}$, we have
 $\la P_1\mu_i,P_1\mu_j\ra=\la\mu_i,\mu_j\ra=\la\hat{\mu}_i,\hat{\mu}_j\ra=\la\hat{P}_1\hat{\mu}_i,\hat{P}_1\hat{\mu}_j\ra$, and
 $\la P_2\nu_i,P_2\nu_j\ra=\la\hat{P}_2\hat{\nu}_i,\hat{P}_2\hat{\nu}_j\ra$. Denote $P_1T_1=\left(\ba{ccc}a_1&b_1&c_1\ea\right)^t$, $P_2T_2=\left(\ba{ccc}a_2&b_2&c_2\ea\right)^t$.
By using $\la P_1\mu_1,P_1\mu_j\ra=\la \hat{P}_1\hat{\mu}_1,\hat{P}_1\hat{\mu}_j\ra$, $j=1,3,5$, i.e.
  $\la P_1T_1,D^rP_1T_1\ra=\la\hat{P}_1\hat{T}_1,D^r\hat{P}_1\hat{T}_1\ra$, $r=0,2,4$, we get
 \be\label{com1}
 t_1^ja_1^2+t_2^jb_1^2+t_3^jc_1^2=t_1^j\hat{a}_1^2+t_2^j\hat{b}_1^2+t_3^j\hat{c}_1^2,~~~ j=0,2,4.
 \ee
Similarly, using $\la P_2\nu_1,P_2\nu_j\ra=\la \hat{P}_2\hat{\nu}_1,\hat{P}_2\hat{\nu}_j\ra$, $j=1,3,5$, and $\la P_1\mu_1,P_1\mu_j\ra=\la \hat{P}_1\hat{\mu}_1,\hat{P}_1\hat{\mu}_j\ra$, $j=2,4,6$, we obtain
\be\label{com2}
t_1^ja_2^2+t_2^jb_2^2+t_3^jc_2^2=t_1^j\hat{a}_2^2+t_2^j\hat{b}_2^2+t_3^j\hat{c}_2^2,~~~ j=0,2,4.\ee
\be\label{com3}t_1^ja_1a_2+t_2^jb_1b_2+t_3^jc_1c_2=t_1^j\hat{a}_1\hat{a}_2+t_2^j\hat{b}_1\hat{b}_2+t_3^j\hat{c}_1\hat{c}_2, ~~~j=1,3,5.
\ee

\begin{enumerate}
\item[1.] If $t_1,t_2,t_3$ are all not equal, from (\ref{com1}) and (\ref{com2}) we can conclude that $\al_i=\pm\hat{\al}_i$ for $\al=a,b,c$ and $i=1,2$.
 \begin{enumerate}
 \item[(i)] If $t_i\neq 0$, $i=1,2,3$, from (\ref{com3}) we get $\al_1\al_2=\hat{\al}_1\hat{\al}_2$ for $\al=a,b,c$. Now if $\al_1\al_2\neq 0$, then we have  $\al_1=\hat{\al}_1\Leftrightarrow \al_2=\hat{\al}_2$. If $\al_1\al_2=0$, suppose $\al_1=0$, then we have $\hat{\al}_1=0$. If $\al_2=\hat{\al}_2$, we also can write $\al_1=\hat{\al}_1$. Let $R=\mathrm{diag}\{e_1,e_2,e_3\}$, where $e_i$ take values $+1$ or $-1$, such that $RP_1T_1=\hat{P}_1\hat{T}_1$. Then one must have $RP_2T_2=\hat{P}_2\hat{T}_2$. Note that the equality (\ref{sd}) is also true if one replaces $P_i$ by $RP_i$. Let $O_1=\hat{P}_1^tRP_1$, $O_2=\hat{P}_2^tRP_2$. We have $\hat{T}_i=O_iT_i$ for $i=1,2$, and $\hat{T}_{12}=O_1T_{12}O_2^t$. To assure that $O_i$ be special, we have $\det R=1$. Firstly, from $\dim\la P_iS_i\ra=\dim\la S_i\ra<3$, we have that $P_iT_i, D^2P_iT_i, D^4P_iT_i$ are linearly dependent. Then there is at least one $\alpha_i^0\in \{a_i,b_i,c_i\}$ that is zero. Hence if $P_1T_1$ and $D^2P_1T_1$ are linearly independent, we have that $DP_2T_2$ can be linearly represented by $P_1T_1$ and $D^2P_1T_2$. Using $t_1t_2t_3\neq0$ and supposing $a_1=0$, we get that $a_2$ is also zero. Now $e_1$ in $R$ can be chosen to be 1 or -1 freely. We can choose $e_1$ to assure that $\det R=1$. Similarly, for the case that $P_2T_2$ and $DP_2T_2$ are linear independent, we can also find $R$ which has determinate one. Lastly, if $P_iT_i$ and $D^2P_iT_i$ are linear dependent, then there are at least two members are zero in $\{a_i,b_i,c_i\}$, $i=1,2$. Therefore, there is an $\alpha\in \{a,b,c\}$ satisfying $\alpha_1=\alpha_2=0$, such that
     $\det R=1$.
 \item[(ii)] If there exists a $t_i=0$, say, $t_3=0$, then we have $\al_1\al_2=\hat{\al}_1\hat{\al}_2$ for $\al=a,b$ from (\ref{com3}). And the invariant $\mathbf{I}$ can assure that $c_1c_2=\hat{c}_1\hat{c}_2$. From the discussion above, we have the conclusion.
 \end{enumerate}
 \item[2.] If there are two different values of $t_1,t_2,t_3$, suppose $t_1=t_2\neq t_3$. Then from (\ref{com1}) and (\ref{com2}), we can get $a_i^2+b_i^2=\hat{a}_i^2+\hat{b}_i^2$, $c_i=\pm \hat{c}_i$ for $i=1,2$.
  \begin{enumerate}
     \item[(i)] If $t_i\neq 0$, $i=1,2,3$, from (\ref{com3}) we get $a_1a_2+b_1b_2=\hat{a}_1\hat{a}_2+\hat{b}_1\hat{b}_2$, $c_1c_2=\hat{c}_1\hat{c}_2$.
     Then there exists a matrix $M\in O(2)$ such that $M\left(\ba{c}a_i\\ b_i\ea\right)=\left(\ba{c}\hat{a}_i\\ \hat{b}_i\ea\right)$, $i=1,2$. And there is an $e=1\ \mathrm{or}\ -1$ such that $ec_i=\hat{c}_i$ for $i=1,2$. Therefore letting $R=\left(\ba{cc}M&\\ &e\ea\right)$, one has $RPT_1=\hat{P}\hat{T}_1$ and $RQT_2=\hat{Q}\hat{T}_2$ again. For the speciality of $R$, from the dimension of $\la S_i\ra$, we have $\det\left(\begin{array}{cc}a_1&a_2\\ b_1&b_2\end{array}\right)=0$ or $c_1=c_2=0$. Hence, we can choose suitable $M$ or $e$ to make sure that $R$ is special.
     \item[(ii)] If $t_1=t_2=0$, we only have $c_1c_2=\hat{c}_1\hat{c}_2$. We can get $M_i\in O(2)$ such that $M_i\left(\ba{c}a_i\\ b_i\ea\right)=\left(\ba{c}\hat{a}_i\\ \hat{b}_i\ea\right)$,
         $i=1,2$, and $R_i=\left(\ba{cc}M_i&\\ &e\ea\right)$ to get the result similarly. We can choose suitable $M_i$ for the speciality of $R_i$.
     \item[(iii)] If $t_3=0$, then one has $R_1,R_2$ with the same $M$ but different $e$ to prove the theorem. The speciality for $R_i$ is similar to the case of $t_i\neq 0$.
  \end{enumerate}
 \item[3.] If $t_1=t_2=t_3\neq 0$, from (\ref{com1}), (\ref{com2}) and (\ref{com3}), we get $a_i^2+b_i^2+c_i^2=\hat{a}_i^2+\hat{b}_i^2+\hat{c}_i^2$ for $i=1,2$, and $a_1a_2+b_1b_2+c_1c_2=\hat{a}_1\hat{a}_2+\hat{b}_1\hat{b}_2+\hat{c}_1\hat{c}_2$.
     Then we have $R\in SO(3)$ such that $RP_1T_1=\hat{P}_1\hat{T}_1$ and $RP_2T_2=\hat{Q}_2\hat{T}_2$. Replacing $P_i$ by $RP_i$ in (\ref{sd}) we get the result.
\item[4.] If $t_1=t_2=t_3=0$, we have $a_i^2+b_i^2+c_i^2=\hat{a}_i^2+\hat{b}_i^2+\hat{c}_i^2$ for $i=1,2$. Therefore  one has $R\in SO(3)$ such that
     $RP_iT_i=\hat{P}_i\hat{T}_i$, $i=1,2$. Replacing $P_i$ by $RP_i$ in (\ref{sd}) one gets the result.
 \end{enumerate}
\hfill \rule{1ex}{1ex}

\medskip
\noindent{\bf Proof of Theorem 3}
 For 3-qubit states $\rho$ and $\hat{\rho}$, they are LU equivalent if and only if
 there are $O_i\in SO(3)$, $i=1,2,3$, such that $\hat{T}_i=O_iT_i$, $\hat{T}_{ij}=O_iT_{ij}O_j^t$ and $\hat{T}_{123}=O_1\otimes O_2\otimes O_3T_{123}$.
Suppose $\dim\la S_i\ra=\dim\la \hat{S}_i\ra=3$, for $i=1,2$. By using the given invariants, we have $O_i\in SO(3)$ such that $\hat{\mu}_i=O_1\mu_i$, $\hat{\nu}_i=O_2\nu_i$ and $\hat{\omega}_i=O_3\omega_i$ for $i=1,2,\cdots$, as well as, $\hat{T}_{12}^t\hat{\mu}_i=O_2T_{12}^t\mu_i$,
 $\hat{T}_{13}^t\hat{\mu}_i=O_3T_{13}^t\mu_i$,
 $\hat{T}_{23}^t\hat{\nu}_i=O_3T_{23}^t\nu_i$ and $\hat{T}_{3|12}\hat{\mu}_i\otimes\hat{\nu}_j=O_3T_{3|12}\mu_i\otimes\nu_j$ for $i,j=1,2,\cdots$.
 Suppose $\mu_{i_1}$, $\mu_{i_2}$ and $\mu_{i_3}$ are linear independent. Then $O_2T_{12}^t(\mu_{i_1}\ \mu_{i_2}\ \mu_{i_3})=\hat{T}_{12}^t(\hat{\mu}_{i_1}\  \hat{\mu}_{i_2}\ \hat{\mu}_{i_3})=\hat{T}_{12}^tO_1(\mu_{i_1}\ \mu_{i_2}\ \mu_{i_3})$. Hence we get $O_2T_{12}^t=\hat{T}_{12}^tO_1$, i.e.
  $\hat{T}_{12}=O_1T_{12}O_2^t$. Similarly, we have $\hat{T}_{13}=O_1T_{13}O_2^t$, $\hat{T}_{23}=O_2T_{23}O_3^t$. From  $\hat{T}_{3|12}\hat{\mu}_i\otimes\hat{\nu}_j=O_3T_{3|12}\mu_i\otimes\nu_j$, $i,j=1,2,\cdots$, we have
  $$\hat{T}_{3|12}O_1\otimes O_2 (\mu_{i_1}\ \mu_{i_2}\ \mu_{i_3})\otimes (\nu_{j_1}\ \nu_{j_2}\ \nu_{j_3})=O_3T_{3|12}(\mu_{i_1}\ \mu_{i_2}\ \mu_{i_3})\otimes (\nu_{j_1}\ \nu_{j_2}\ \nu_{j_3}),$$
  where $\nu_{j_1},\nu_{j_2},\nu_{j_3}$ are linear independent vectors in $S_2$. Using the linear independence of $\mu_{i_1},\mu_{i_2},\mu_{i_3}$ and  $\nu_{j_1},\nu_{j_2},\nu_{j_3}$, we get $\hat{T}_{3|12}O_1\otimes O_2=O_3T_{3|12}$ or
  $\hat{T}_{3|12}=O_3T_{3|12}O_1^t\otimes O_2^t$ which is equivalent to $\hat{T}_{123}=O_1\otimes O_2\otimes O_3T_{123}$.
\hfill \rule{1ex}{1ex}

\newpage
\bigskip
\noindent{\bf Acknowledgments}\, \, We would like to thank the referee for pointing out a
serious mistake in the initial draft of the paper.
This work is supported by NSFC (Grant No. 11401339, 11275131, 11675113), NSF of Shandong (No. ZR2014AQ027).

\bigskip
\noindent{\sf Author contributions}

\noindent  B.-Z.S., S.-M.F. and Z.-X.W. wrote the main manuscript text. All of the authors reviewed the manuscript.

\bigskip
\noindent{\sf Additional Information}

\noindent Competing Financial Interests: The authors declare no competing financial interests.

\end{document}